\documentclass[letterpaper, 10 pt, conference]{ieeeconf} 

\IEEEoverridecommandlockouts    

\usepackage{cite}
\usepackage{amsmath,amssymb,amsfonts}
\usepackage{algorithmic}
\usepackage{graphicx}
\usepackage{textcomp}
\usepackage{xcolor}
\usepackage{listings}
\usepackage{url}
\usepackage{tikz}
\usepackage{subcaption}
\usepackage{mathalfa}
\usepackage{hyperref}
\let\labelindent\relax

\usepackage[inline]{enumitem} 
\newtheorem{theorem}{Theorem}[section]
\newtheorem{remark}{Remark}[section]
\newtheorem{property}{Property}[section]

\DeclareUnicodeCharacter{2194}{\ensuremath{\leftrightarrow}}

\let\hm=\mathcal
\lstdefinestyle{matlabstyle}{
 language=Matlab,
 basicstyle=\scriptsize\ttfamily,
 keywordstyle=\color{blue},
 commentstyle=\color{green!40!black},
 stringstyle=\color{red},
 numberstyle=\tiny\color{gray},
 breaklines=true,
 frame=single,
 backgroundcolor=\color{gray!5}
}

\lstdefinestyle{interpretercode}{
 basicstyle=\scriptsize\ttfamily,
 numberstyle=\tiny\color{gray},
 breaklines=true,
 frame=leftline,
 framerule=2pt,
 rulecolor=\color{blue!30},
 backgroundcolor=\color{gray!5}
}

\title{\LARGE \bf
The PhasorArray Toolbox for Harmonic Analysis and Control Design
}

\author{Maxime Grosso$^{1}$, Pierre Riedinger$^{1}$ and Jamal Daafouz$^{1, 2}$
\thanks{$^{1}$M. Grosso, P. Riedinger and J. Daafouz are with Univ. Lorraine, CNRS, CRAN, 
 Nancy, France. (e-mail: maxime.grosso@univ-lorraine.fr; pierre.riedinger@univ-lorraine.fr; jamal.daafouz@univ-lorraine.fr)}%
\thanks{$^{2}$J. Daafouz is also with Insitut Universitaie de France, Paris, France.}%
}

\begin{document}

\maketitle
\thispagestyle{empty}
\pagestyle{empty}

\begin{abstract}
We present a MATLAB package called the PhasorArray Toolbox that has been developed to make harmonic analysis and control methods both practical and user-friendly. The toolbox adopts an object-oriented architecture that enables intuitive manipulation of periodic matrices through overloaded operators for addition, multiplication, convolution, and automatic Toeplitz construction. Its advanced features include harmonic Sylvester, Lyapunov and Riccati equations solvers, and seamless integration with YALMIP, thereby facilitating advanced control and analysis techniques based on Linear Matrix Inequalities (LMIs) in the harmonic framework.
\end{abstract}

\section{Introduction}
In control applications involving periodic or oscillatory dynamics~\cite{farkas_periodic_1994}, harmonic domain analysis offers a powerful framework for modeling and controller design~\cite{Blin2022NecessaryTime,salis_stability_2017,karami_dynamic_2018,lyu_harmonic_2019}. For time-periodic (TP) systems~\cite{bittanti_periodic_2009}, this approach yields an equivalent infinite-dimensional time-invariant (TI) representation in the harmonic domain~\cite{Blin2022NecessaryTime}. Numerical algorithms are available for solving infinite-dimensional harmonic LMIs, as well as algebraic Lyapunov and Riccati equations, to arbitrarily high precision (e.g.,~\cite{riedinger_solving_2022,vernerey_tblmi_2025,grosso_harmonic_2025}). These tools enable direct controller synthesis in the infinite-dimensional setting. Regarding numerical tools, to the best of our knowledge, only a few existing implementations focus primarily on lifting-based methods dedicated to linear time-periodic (LTP) systems and lack LMI capabilities \cite{Varga}. Consequently, there remains a clear need for accessible MATLAB-based tools specifically tailored to harmonic domain analysis. The practical implementation of harmonic methods involves intricate manipulations of bi-infinite Toeplitz matrices, careful handling of truncation effects, and dedicated algorithms for Lyapunov, Riccati, and LMI formulations.


The proposed PhasorArray Toolbox, openly available on GitHub \cite{PhasorArray25}, overcomes these implementation challenges by systematically encapsulating harmonic operations within an object-oriented MATLAB framework. Built around a core \texttt{PhasorArray} class, the toolbox provides a structured representation for storing phasors associated with $T$-periodic matrices in three-dimensional arrays, with algebraic operations seamlessly supported through operator overloading. The toolbox addresses key challenges in harmonic modeling, analysis, and control by providing comprehensive tools for: \begin{enumerate*}[label=(\roman*)] \item constructing \texttt{PhasorArray} objects from $T$-periodic matrices and reconstructing time-domain representations, \item performing consistent arithmetic operation ($+,\times,\cdot ^{-1}$) on \texttt{PhasorArray}, \item visualizing harmonic-domain and time-domain data, \item applying truncation techniques for model simplification, \item evaluating system properties, \item solving harmonic Riccati, Sylvester, and Lyapunov equations, and \item formulating  Toeplitz-Block LMI problems for robust control design\end{enumerate*}. 

The paper is organized as follows. Section II revisits the harmonic framework and its numerical challenges. Section III presents the toolbox architecture and capabilities, while Section IV illustrates basic construction, manipulation, and visualization using a simple periodic matrix example. Section V focuses on applications to harmonic analysis and robust control. Although an LTP system is used here for illustration, the framework extends naturally to more complex periodic systems. The paper concludes by summarizing the main contributions and benefits of the toolbox.
\section{Harmonic Modeling and Implementation Challenges}

Before introducing the toolbox, we first revisit the harmonic framework and outline the associated numerical challenges.

For any scalar time-signal $x\in L^2_{loc}(\mathbb{R},\mathbb{R})$ and a given period $T = 2\pi/\omega$, we define the \textit{phasors} $X=\hm F(x)$ associated to $x$, by the time-varying sequence $X(t)=(X_k(t))_{k\in\mathbb{Z}}$ whose components result from a sliding Fourier decomposition given by:
\begin{equation}
X_k(t) = \frac{1}{T}\int_{t-T}^{t} x(\tau)e^{-\texttt{j}k\omega \tau} d\tau.
\end{equation}
For a time vector-valued function $x\in L^2_{loc}(\mathbb{R},\mathbb{R}^n)$ (or matrix-valued), this extends to \begin{equation}
 X=\hm F(x)=(\hm F(x_1),\cdots,\hm F(x_n)).\label{fourier}
\end{equation}
The Toeplitz transformation $\hm T$ of a scalar function $a\in L^\infty$ with phasors $({a_k})_{k \in \mathbb{Z}}=\hm F(a)$ is defined by the infinite dimensional Toeplitz operator bounded on $\ell^2$:
 \begin{equation}\scriptsize\label{eq:toeplitz_matrix_def}
 \hm{T}(a)=
 \begin{bmatrix}
 \ddots & \vdots & \vdots & \vdots& \\
 \cdots & a_{0} & a_{-1} & a_{-2} & \cdots \\
 \cdots & a_{1} & a_{0} & a_{-1} & \cdots \\
 \cdots & a_{2} & a_{1} & a_{0} & \cdots \\
 & \vdots & \vdots & \vdots & \ddots
 \end{bmatrix}.
\end{equation}
A matrix-valued function $A={(a_{ij})}_{i\in\{1,n\}, j\in\{1,m\}}$, is represented by the  Toeplitz-Block (TB) operator 
 
\begin{equation}
 \hm A = {(\hm T(a_{ij}))}_{i\in\{1,n\}, j\in\{1,m\}}. \label{toeplitz}
\end{equation}
We recall the following algebraic properties:
\begin{property}\label{prop:product}
For a signal $x\in L^2_{\text{loc}}$ and $L^\infty$-matrix valued function $A$ and $B$:
\begin{align}
 \hm{F}(Ax)& = \hm{T}(A)\hm{F}(x) = \hm{A}X, \label{e1}\\
 \hm{T}(AB) &= \hm{T}(A)\hm{T}(B) = \hm{A}\hm{B}.\label{e2}
 \end{align}
 \end{property}
We have the following fundamental result:
\begin{theorem}\label{equiv_mod}Consider a general dynamical system \begin{equation}
 \dot{x}(t) = f(t, x(t)),\label{st}
\end{equation} admitting Caratheodory's solution. Then, for any $T>0$, $X=\hm F(x)$ satisfies:
\begin{align}\label{eq:harmodel_fixed_freq}
 \dot{X}(t) =& \hm{F}(f(t, x(t))) - \hm{N} X(t),
\end{align}
where $\hm{N} = \mathrm{Id}_n \otimes \mathrm{diag}(	\texttt{j} \omega \mathbb{Z})$ and conversely if $X$ belongs to $\hm H=\textrm{Im}_{\hm F}(L^2_{loc})$ (see \cite{Blin2022NecessaryTime} for a characterization of $\hm H$). 
\end{theorem}

This theorem is particularly useful when analyzing \(T\)-periodic systems (i.e., \(f(t,x)=f(t+T,x)\)), since in that case their harmonic representation~\eqref{eq:harmodel_fixed_freq} becomes an infinite-dimensional \emph{time-invariant} system. Thus, methods designed for time-invariant systems can be reformulated and applied in this infinite-dimensional framework, as long as the system’s infinite dimensionality is appropriately considered.

However, defining the phasor sequence from time-domain signals, constructing the associated  Toeplitz-Block operator, and performing the necessary algebraic manipulations—such as those given in~\eqref{e1} and~\eqref{e2}—typically involve extensive and cumbersome manual coding.
Moreover, due to the infinite-dimensional nature of harmonic representations, \emph{truncation} is essential for numerical computations. The truncation scheme used must be \emph{consistent}, in the sense that it allows recovery of the infinite-dimensional solution as the truncation order increases. In practice, however, properties such as~\eqref{e1} and~\eqref{e2} no longer hold for truncated operators, leading to inconsistencies and potential convergence issues. Specifically,
\begin{equation}\label{eq:noncommute_trunc}
 \mathcal{T}(AB)_h \neq \mathcal{T}(A)_h \, \mathcal{T}(B)_h,
\end{equation}
where the subscript \(h\) denotes truncation to order \(h\) of the  Toeplitz-Block matrices, corresponding to the extraction of the principal sub-matrix of dimension \((2h+1)\times(2h+1)\) from each \(n\times m\) block, with \(n\times m\) referring to the size of \(A\) in the time domain.

As a result, attempting to solve a Lyapunov equation using the truncated version of the Toeplitz–block operator may fail to correctly establish the stability of the corresponding system (see \cite{riedinger_solving_2022} for a complete explanation).

\section{PhasorArray Toolbox}
\subsection{Design Philosophy and Architecture}

The toolbox design centers on systematic storage of periodic matrices using 3D arrays with overloaded operators. Any periodic matrix $A(t)$ with period $T$ is expressed as a \textit{finite truncated sum}:
\begin{equation}
A(t) = \sum_{k=-h}^{h} A_k e^{\texttt{j}k\omega t},
\end{equation}
where the harmonic coefficients $A_k$ capture the spectral content \footnote{It is assumed that the remainder 
$
\|\sum_{|k|>h} A_k e^{\texttt{j}k\omega t}\|_{L^\infty} \to 0$ as $h \to \infty$.
If this almost everywhere uniform convergence condition is not satisfied, one may apply a regularization technique—such as convolution with a smooth bump function—to ensure this property holds.
}($h$ is assumed sufficiently large). 

A \texttt{PhasorArray} object encapsulates this representation by \textbf{storing the coefficients $A_k$ as slices of a 3D array}, with the DC component $A_0$ always positioned on the \textbf{central slice}.
This structured storage approach enables the natural use of MATLAB syntax for complex harmonic operations while ensuring that all subsequent computations are carried out systematically and efficiently. 

\begin{figure}[ht!]
\centering
\begin{tikzpicture}
\draw[thick] (0,0) rectangle (2,1.5);
\draw[thick] (0.3,0.3) rectangle (2.3,1.8);
\draw[thick] (0.6,0.6) rectangle (2.6,2.1);

\node at (1.3+0.9,-0.3) {$A_{-h}$};
\node at (1.6+0.9,0) {$A_{-1}$};
\node at (1.9+0.9+0.5,0.3) {$A_0$ (center)};
\node at (2.2+0.9,0.6) {$A_{+1}$};
\node at (2.5+0.9,0.9) {$A_{+h}$};

\draw[->,thick] (1.3+1.7+1,-0.3) -- (2.5+1.7+1,0.9) node[right] {slice index $k$};
\end{tikzpicture}
\caption{3D array structure with harmonic coefficients stored as slices: $\texttt{A(:,:,h+1)}$ is the DC component, $\texttt{A(:,:,h+2)}$ is the 1rst harmonic, $\texttt{A(:,:,1)}$ is the $-h$th harmonic.}
\end{figure}
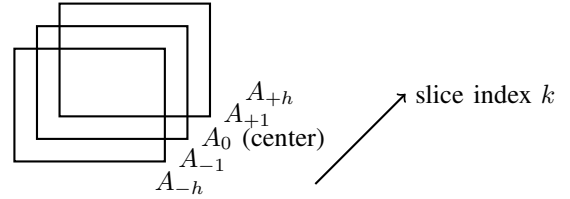

{More precisely, the toolbox provides the following functionalities:
\begin{itemize}
 \item Natural arithmetic operations (addition, multiplication) and convolution implemented via operator overloading.
 \item Construction of truncated operators: \texttt{T\_tb}  Toeplitz-Block Matrix ($\hm T(A)$), \texttt{N\_tb} ($\hm N$) and \texttt{F\_tb} ($\hm F(A)$)\footnote{\texttt{\_tb} denotes the compatibility with Toeplitz-Block structure, alternatively a Block-Toeplitz equivalent structure is obtained using \texttt{\_bt}, see remark~1 p. 2539 in\cite{vernerey_tblmi_2025} and \cite{Blin2022NecessaryTime}.}. 
 \item Visualization and diagnostic tools (plot, stem, barsurf).
 \item Solver and synthesis utilities (Lyapunov/Sylvester, iterative Riccati, \texttt{PhasorSS}) and YALMIP interoperability~\cite{lofbergYALMIPToolboxModeling2004}.
\end{itemize}
These components help keep examples concise and focused on modeling and control aspects rather than low-level indexing details.}

\subsection{Extended capabilities}


Beyond its core functionalities, the toolbox also provides utilities for harmonic extraction (\texttt{time2Phasor}, \texttt{funcToPhasorArray}), test-case generation (\texttt{PhasorArray.random}), model reduction (\texttt{neglect}, \texttt{trunc}), visualization (\texttt{stem}, \texttt{barsurf}), time and phase evaluation (\texttt{evalp}, \texttt{evaltime}), and simulation of periodic state-space systems (\texttt{PhasorSS}). Advanced features include YALMIP integration (\texttt{ndsdpvar}) for LMI formulation and dynamic phasor computation (\texttt{angularsft}).

\section{Basic construction, manipulation and visualization}

We illustrate below, through a simple example, how to create and manipulate PhasorArray objects.
\subsection{Building a PhasorArray from a Periodic Matrix}
Consider a $2\times 2$ periodic matrix with period $T=1$:
\begin{equation} 
A(t) = 
\begin{bmatrix}
\frac{3}{2} + \text{sawtooth}(\omega t, 0.5) & 1 + \cos(\omega t) \\ 
1 - \sin(2\omega t) & \frac{\text{square}(\omega t)-1}{2}
\end{bmatrix},\label{ex}
\end{equation}
whose complex Fourier coefficients $A_{ij,k}$, for  $i,j=1,2$ and $k\in \mathbb{Z}$, are given by: 
\begin{align*}
A_{11,0}&=\frac{3}{2},\, A_{11,2k}=0, \, A_{11,2k+1} = \frac{4(-1)^{k+1}}{\pi^2 (2k+1)^2},\\
A_{12}&=(\cdots,0,0,\frac{1}{2},\underline{1},\frac{1}{2},0,0,\cdots),\\
A_{21}&=(\cdots,0,0,\frac{-	\texttt{j}}{2},0,\underline{1},0,\frac{	\texttt{j}}{2},0,0,\cdots),\\
A_{22,0}&=-\frac{1}{2},\, A_{22,2k}=0,\, A_{22,2k+1} = \frac{1}{	\texttt{j}\pi (2k+1)}.
\end{align*}
We now show two ways to build the corresponding \texttt{PhasorArray}: by directly specifying the phasors, or by using a {time periodic function handle.}

\noindent
\textbf{Construction of a PhasorArray by specifying the phasors:}
A direct construction can be achieved by providing all harmonics explicitly (negative to positive) up to a given order or only DC and positive harmonics for real signals ($X_{-k}=\bar X_k$). Let us build a PhasorArray by considering only the first three harmonics as follow:

\begin{lstlisting}[style=matlabstyle]
% Define harmonic coefficients
A_0 = [1.5 1; 1 -0.5];
A_1 = [-4/pi^2 1/2; 0 1/(pi*1i)];
A_2 = [0 0; 1i/2 0];
A_3 = [4/(3*pi)^2 0; 0 1/(pi*3*1i)];

% Create PhasorArray with all harmonics (negative to positive)
A = PhasorArray(cat(3,conj(A_3), conj(A_2), conj(A_1), A_0, A_1, A_2, A_3));
% Equivalently for real signal:
A = PhasorArray(cat(3, A_0, A_1, A_2, A_3),"isreal",true)
\end{lstlisting}

\begin{lstlisting}[style=interpretercode]
>> A =
2x2x7 PhasorArray of double representing a 2x2 real-valued periodic matrix with 3 harmonics
\end{lstlisting}

In this example, the PhasorArray 
$A$ is not an accurate representation of the matrix-valued function 
$A(t)$, since only a very limited number of harmonics were considered. To overcome this issue, it is preferable to proceed with the following alternative.\\
\noindent
\textbf{Construction of a PhasorArray using a function handle:} The PhasorArray $A$ can be built by sampling the time-domain signal on a $2^N$ grid over one or several periods and extracting harmonics via FFT. Note that if one period is used, the number of harmonics computed is now $h=2^{N-1}-1$.
\begin{lstlisting}[style=matlabstyle]
T = 1; %period
N = 6; %2^N points used for FFT 
At = @(t) [1+sawtooth(2*pi*t/T,0.5)+0.5, 1+cos(2*pi*t/T); 1-sin(2*2*pi*t/T), -0.5 + square(2*pi*t/T)/2];
A = PhasorArray.funcToPhasorArray(At,T,N)
\end{lstlisting}
\begin{lstlisting}[style=interpretercode]
>> A = 
2x2x63 PhasorArray of double representing a 2x2 real-valued periodic matrix with 31 harmonics
\end{lstlisting}

\subsection{Visualization Methods}
\begin{figure}[!ht]
\centering
\begin{subfigure}{0.49\columnwidth}
\centering
\includegraphics[width=\linewidth]{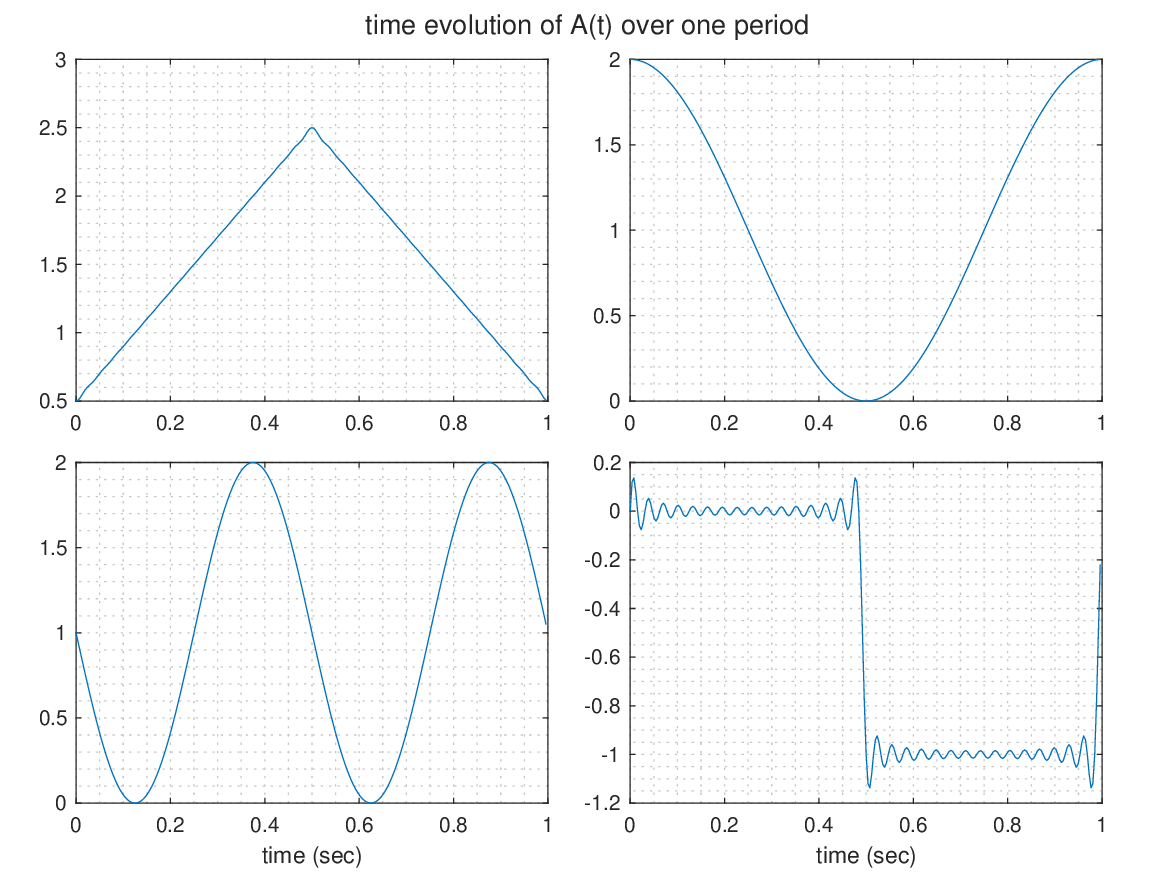}
\caption{Time evolution of $A(t)$ with \texttt{plot(A)}.}
\label{fig:time_evolution_A}
\end{subfigure}\hfill
\begin{subfigure}{0.49\columnwidth}
\centering
\includegraphics[width=\linewidth]{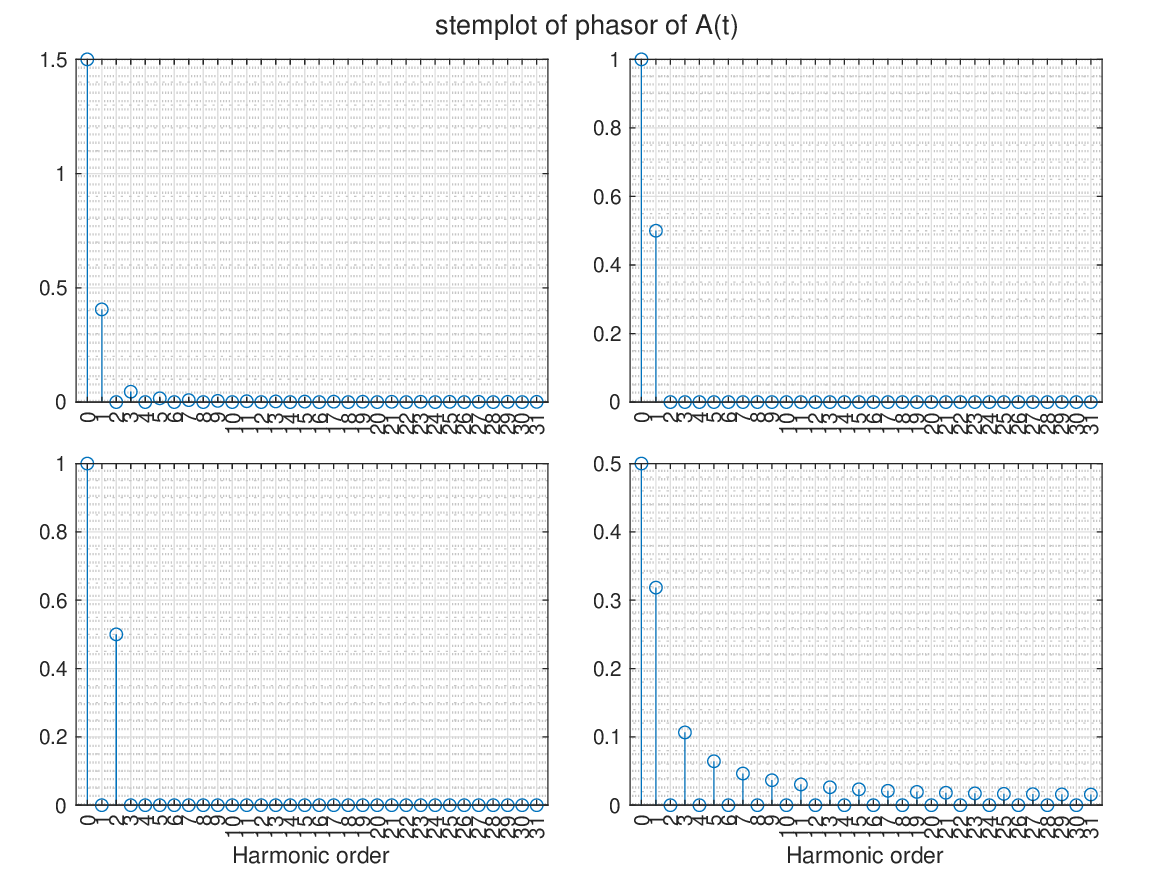}
\caption{Harmonic spectrum with \texttt{stem} plot. }
\label{fig:stem_A}
\end{subfigure}
\caption{Visualization examples for periodic matrix $A(t)$ approximated by $h=31$ harmonics. }
\end{figure}
The toolbox provides multiple visualization modes, as depicted in figures~\ref{fig:time_evolution_A} and~\ref{fig:stem_A}.

\begin{lstlisting}[style=matlabstyle]
% Harmonic spectrum visualization (stem plot)
figure 
stem(A,"scale","linear","uniformYLim",0)
sgtitle('stemplot of phasor of A(t)')

% Time-domain reconstruction over multiple periods
figure
plot(A)
sgtitle('time evolution of A(t) over one period')
\end{lstlisting}

Figure~\ref{fig:time_evolution_A} shows the time-domain realization by evaluating $A(t) = \sum_{k=-h}^h A_k e^{\texttt{j}k\omega t}$ while the \texttt{stem} plot (figure~\ref{fig:stem_A}) displays harmonic magnitudes across all matrix elements.

\subsection{Harmonic reduction} The toolbox includes methods for harmonic reduction via thresholding or fixed-order truncation, along with visualization of the impact on time-domain reconstruction. Furthermore, the brackets operator \texttt{\{\}} allows direct access to the underlying phasorArray at given coordinates.

\begin{lstlisting}[style=interpretercode]
>> A{2,2}
1x1x63 PhasorArray of double representing a 1x1 real-valued periodic matrix with 31 harmonics
\end{lstlisting}

\begin{lstlisting}[style=matlabstyle]
% Neglect small harmonics (threshold-based filtering)
A_neglect = neglect(A{2,2}, 2.e-2, 'reduceMethod', 'absolute');
% Truncate to fixed harmonic order
A_trunc = trunc(A{2,2}, 5);
figure
stem(A{2,2}, 'scale', 'log')
hold on
stem(A_neglect, 'Marker', 'x')
stem(A_trunc, 'Marker', '<')
legend('PhasorArray','Neglected Phasors','Truncated Phasors')
title('Harmonic content (log scale)')
xlabel('Harmonic'), ylabel('Magnitude')
figure
plot(A{2,2}, T, [0, 2*T])
hold on
plot(A_neglect, T, [0, 2*T])
plot(A_trunc, T, [0, 2*T])
t=0:.01:2*T;
A_eval = arrayfun(@(t) At(t), t, 'UniformOutput', false);
catA_eval = cat(3, A_eval{:});
plot(t, squeeze(catA_eval(2,2,:)), 'k--')
legend('PhasorArray', 'Neglected Phasors', 'Truncated Phasors', 'Original Signal')
title('Square wave'), xlabel('Time (s)'), ylabel('Amplitude')
\end{lstlisting}

\begin{figure}[!ht]
\centering
\begin{subfigure}{0.49\columnwidth}
\centering
\includegraphics[width=\linewidth]{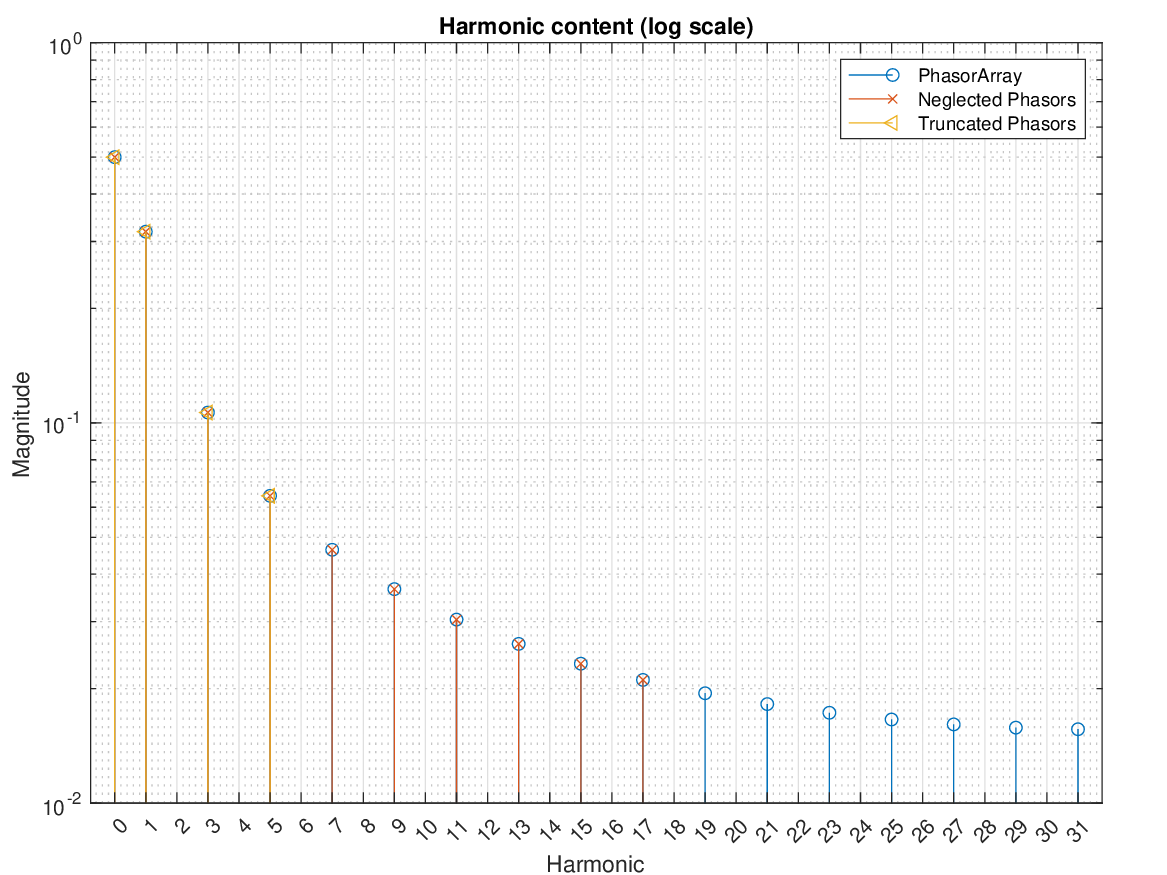}
\caption{Stem }
\label{fig:triangular_wave_a}
\end{subfigure}
\hfill
\begin{subfigure}{0.49\columnwidth}
\centering
\includegraphics[width=\linewidth]{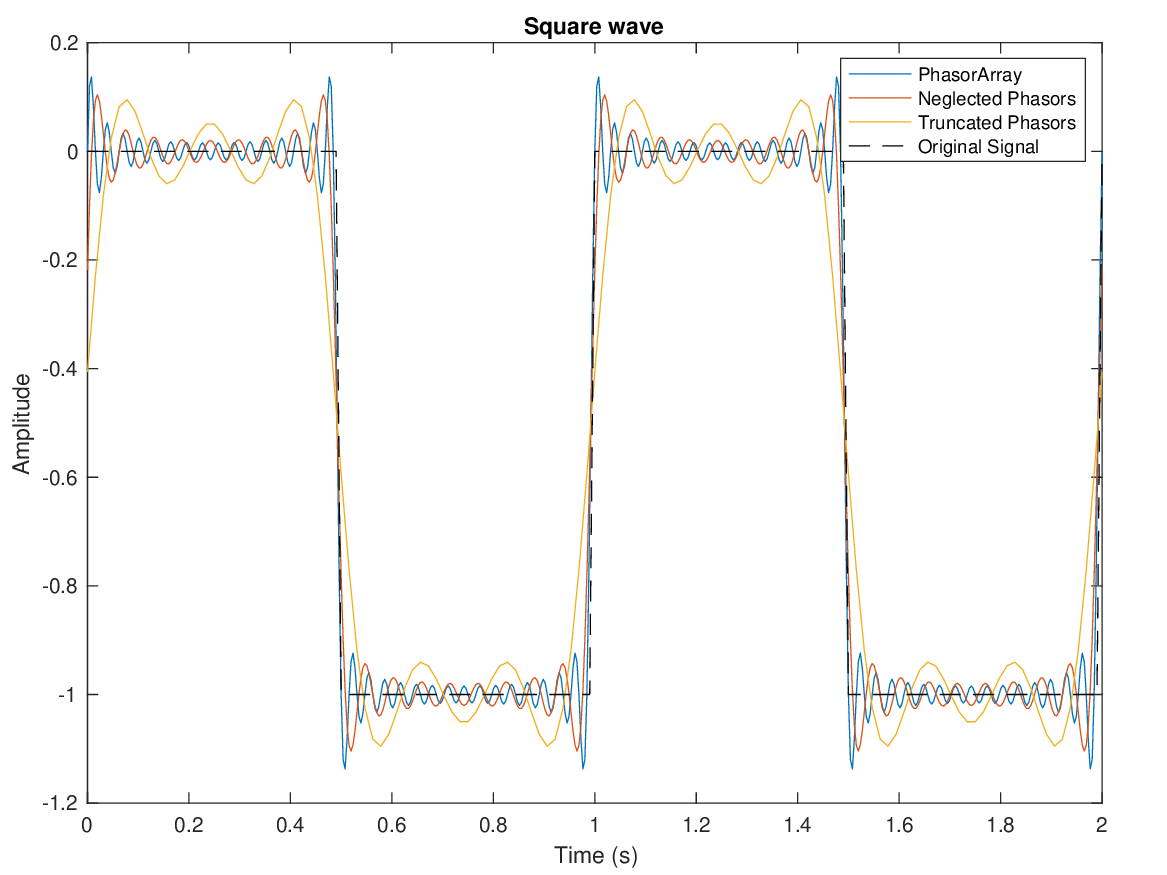}
\caption{time-domain}
\label{fig:triangular_wave_b}
\end{subfigure}
\caption{Visualization example. Left: harmonic spectrum showing full representation, threshold-based filtering (\texttt{neglect}), and fixed-order truncation (\texttt{trunc}). Right: time-domain reconstruction comparing accuracy of different reduction methods.}
\label{fig:visu_subfig}
\end{figure}

The \texttt{neglect()} function removes harmonics below a threshold (adaptive reduction), while \texttt{trunc()} enforces a fixed harmonic order, demonstrating trade-offs between accuracy and model complexity.

\subsection{Basic PhasorArray Operations}
Standard MATLAB operations have been overloaded to enable intuitive and efficient manipulation of PhasorArrays.
\textit{Key insights:} 
\begin{itemize} 
 \item Temporal multiplication \texttt{D = A * B} performs harmonic convolution, implementing $D_k = \sum_{i+j=k} A_i B_j$.
 \item Inversion \texttt{inv(A)} computes the pointwise inverse $A(t)^{-1}$ in the time domain and reconstructs the harmonic components via FFT.
 \item Transpose and Hermitian conjugate are defined with respect to the time-domain representation $A(t)$, operating on the corresponding harmonic slices.
\end{itemize}

\begin{lstlisting}[style=matlabstyle]
% Generate random periodic matrix
B = PhasorArray.random(2, 2, 2)
% 2x2 with 2 harmonics

% Algebraic operations (computed in harmonic domain)
   % Equivalence in the time domain
C = A + B;  % Addition: C(t) = A(t) + B(t)
D = A * B;  % Multiplication: D(t) = A(t)*B(t)
Ainv = inv(A); % Inversion: Ainv(t) = A(t)^{-1}
At = A.';  % Transpose
Ah = A';  % Transpose conjugate
\end{lstlisting}

\subsection{Finite dimensional Fourier or  Toeplitz-Block extraction representations from a PhasorArray}

 The Toeplitz operator $\mathcal{T}$ maps a periodic matrix $A(t)$ to an infinite  Toeplitz-Block matrix $\mathcal{A} = \mathcal{T}(A)$. For practical implementation, a finite truncated representation up to harmonic order $h$ is used. If $A(t)$ is of dimension $n\times m$, $\mathcal{T}(A)_h$ is formed by $n\times m$ blocks of Toeplitz matrices of dimension $(2h+1) \times (2h+1)$. From the PhasorArray representation, the construction proceeds as follows:

\begin{lstlisting}[style=matlabstyle]
% Create  Toeplitz-Block form truncated to h=8 harmonics
h = 8;
A_tb = A.T_tb(h); % Toeplitz-Block form (full matrix)
% Visualize the structure
figure
barsurf(abs(A_tb),1e-3); 
title('T(A)_8');
\end{lstlisting}

\begin{figure}[ht!]
\centering
\includegraphics[width=0.8\columnwidth]{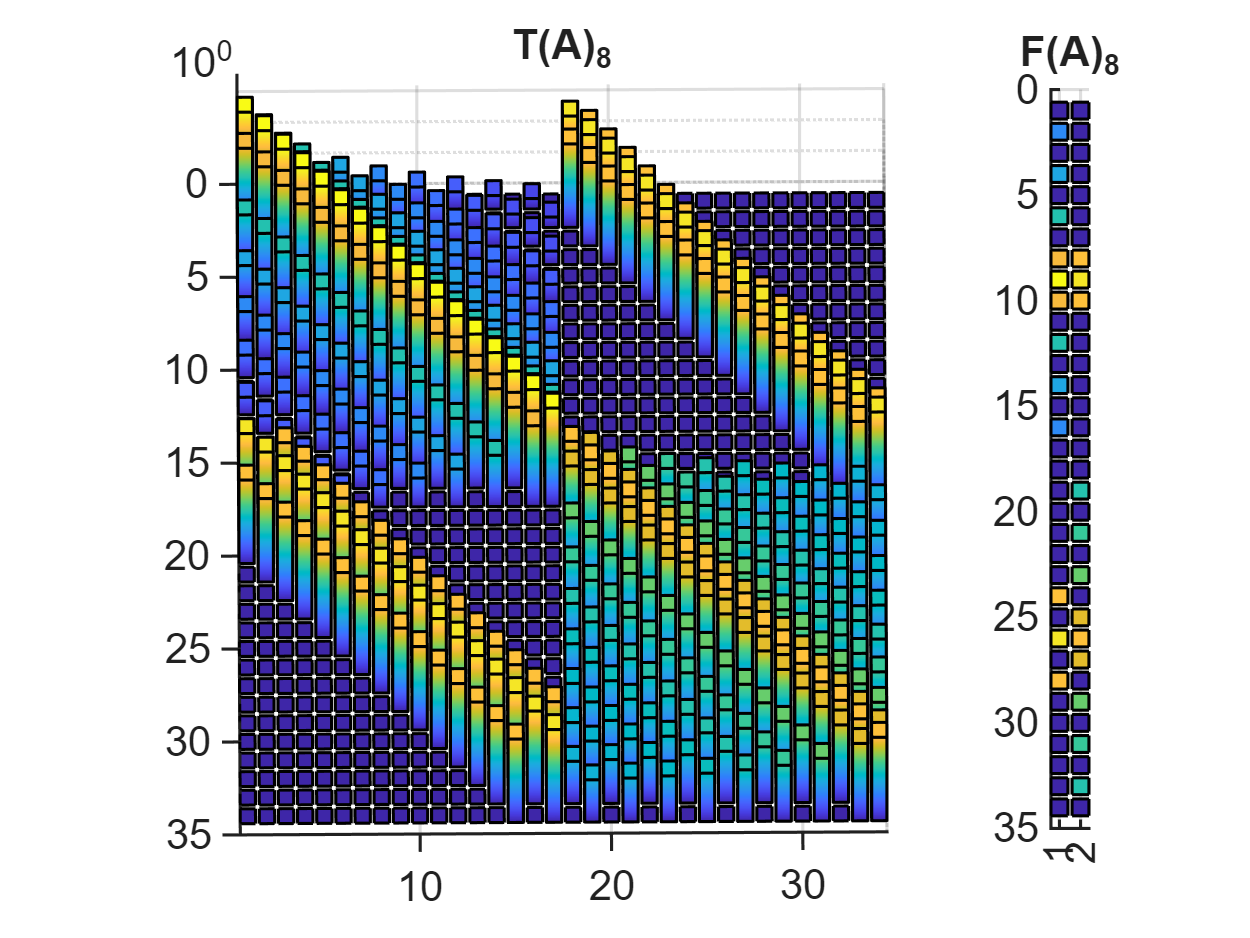}
\caption{Barsurf of absolute value of $\mathcal{T}(A)_h$ of $((2h+1)n\times (2h+1)m)$ dimension and $\mathcal{F}(A)_h$ of $(2h+1)n\times m)$ dimension for $h=8$.}
\label{fig:toeplitz_spy}
\end{figure}
In the same spirit, one can obtain from a PhasorArray $A$, its truncated Fourier representation (see equation \eqref{fourier}), $\hm{F}(A)_h$ as a $(2h+1)n \times m$ matrix. Both representations are illustrated in Figure~\ref{fig:toeplitz_spy}.

\begin{lstlisting}[style=matlabstyle]
A_four = F_tb(A,h) % Fourier form of PhasorArray A 
\end{lstlisting}

\section{Control design for LTP Systems}

With the construction, manipulation, and visualization of the PhasorArray established, we next demonstrate its application to harmonic analysis and control design. Although an LTP system is used here as an illustrative example due to space constraints, the framework naturally extends to more complex periodic systems, including periodic bilinear, power, and drive systems, for both control design and harmonic mitigation \cite{Nwaneto,grosso_harmonic_2025}.

\subsection{Linear Time-Periodic Systems}
As an example, we consider the following LTP system:
\begin{equation}
\dot{x}(t) = A(t)x(t) + B(t)u(t),
\end{equation}
with $T$-periodic matrix-valued function $A$ and $B$ belonging to $L^\infty$. In the harmonic domain, this transforms into:
\begin{equation}
\dot{X}(t) = (\mathcal{A} - \mathcal{N})X(t) + \mathcal{B}U(t),
\end{equation}
where $\mathcal{N} = I_n \otimes 	\texttt{j} \omega\text{diag}(\ldots,-2, -1, 0, 1, 2, \ldots) $ is the harmonic differentiation operator and $\mathcal{A}=\hm T(A)$ and $\mathcal{B}=\hm T(B)$ are bounded on $\ell^2$.
$\hm N$ is obviously not a Toeplitz operator and can be built as follows:
\begin{lstlisting}[style=matlabstyle]
% For system with state dimension n=2 and period T=1
T = 1;
h = 10; % Truncation order
% Construct N operator in Toeplitz-Block form
Ntb = N_tb(size(A,1), h, T); 
\end{lstlisting}

To perform a stability analysis, the eigenvalues of $\mathcal{A} - \mathcal{N}$ are the Floquet exponents of the periodic system and are given by $\sigma{(\hm{A -N})} = \{\lambda_i + 	\texttt{j} k \omega\,; i=1,\cdots,n; k \in\mathbb{Z} \}$ where the $\lambda_i$ can be computed as follows: 

\begin{lstlisting}[style=matlabstyle]
lambda = A.HmqNEig(h,T,"fundamental")
\end{lstlisting}
\begin{lstlisting}[style=interpretercode]
>> lambda =
   1.9060 + 0.0000i
  -0.9060 + 0.0000i
\end{lstlisting}


The analysis clearly shows that the LTP system is unstable

\subsection{Lyapunov Equations}
For a periodic system $\dot{x} = A(t)x$, asymptotics stability can also be analyzed via the existence of a $T-$periodic, symmetric and positive definite solution to the periodic differential Lyapunov equation:
\begin{equation}\small
\dot{P}(t) + A(t)^T P(t) + P(t) A(t) + Q(t) = 0,\ P(0)=P(T),
\end{equation}
where $Q(t)=Q(t)'>0$ is $T-$periodic and belongs to $L^\infty$.
In the harmonic domain, this is equivalent to solve the harmonic Lyapunov equation:
\begin{equation}\label{eq:lyapHm}
(\mathcal{A}- \mathcal{N})^* \mathcal{P} +\mathcal{P} (\mathcal{A}-\mathcal{N}) + \mathcal{Q} = 0,
\end{equation}
and $(\mathcal{A}- \mathcal{N})$ is Hurwitz if and only if the solution $\hm P$ is hermitian positive definite (and bounded on $\ell^2$). 

A consistent\footnote{Here, consistency refers to the property that when ${A}$ and ${Q}$ are truncated to harmonic order $k$, the corresponding solution $P(k)$ converges to the infinite-dimensional solution ${P}$ as $k\rightarrow+\infty$.} scheme for solving~\eqref{eq:lyapHm} has been proposed in~\cite{riedinger_solving_2022}, leading to the following implementation in the toolbox:

\begin{lstlisting}[style=matlabstyle]
Q = PhasorArray.eye(2); % Positive definite Q(t)
P = lyap(A,Q,"T",T);
\end{lstlisting}

From the resulting PhasorArray $P$, the T-periodic matrix $P(t)$ can be evaluated as follows
\begin{lstlisting}[style=matlabstyle]
t=0:0.1:T;
Pt = evalTime(P,T,t);
\end{lstlisting}

\subsection{Riccati Equations and LQR Control}
Consider the classical LQR problem. The $T$-periodic feedback gain, $K(t) = R(t)^{-1}B(t)^T S(t)$ is computed by solving the periodic differential Riccati equation \cite{colaneri_continuous-time_2000}:
\begin{equation}
\dot{S} + A^T S + S A - S B R^{-1} B^T S + Q = 0,\ S(0)=S(T),
\end{equation}
where it is assumed that all matrix entries are $T$-periodic and belong to $L^\infty$.
In the harmonic domain, the problem reduces to solving the following algebraic Riccati equation:
\begin{equation}
 \hm{(A-N)^* S + S (A-N) - S B R}^{-1}\mathcal{B^* S + Q } = 0.\label{ric}
\end{equation}
The algorithm proposed in~\cite{riedinger_solving_2022} is implemented in the toolbox as \texttt{RicHarmonicKlein()}. It employs an iterative Kleinman-like approach, which guarantees the consistency of the solution. Note that Kleinman’s algorithm requires an initial gain $K_0$ that ensures the stability of the closed loop.

\begin{lstlisting}[style=matlabstyle]
B = [1;PhasorArray.sin]
Q = PhasorArray.eye(2)*10;
R = PhasorArray.eye(1);
K0 = PhasorArray([10,10]);
htrunc = 6; %initial truncation order for Riccati solution

[K_final, S_final] = RicHarmonicKlein(A, B, Q, R, K0,T,"autoUpdateh",true,"max_iter",50,"residualThreshold",1e-6,"htrunc",htrunc,"hmax",500)
\end{lstlisting}

\begin{lstlisting}[style=interpretercode]
Converged at iteration 9
Riccati residual norm: 2.37e-07

K_final = 
1x2x157 PhasorArray of double representing a 1x2 real-valued periodic matrix with 78 harmonics

S_final = 
2x2x155 PhasorArray of double representing a 2x2 real-valued periodic matrix with 77 harmonics
\end{lstlisting}

The origin of the system is asymptotically stable in closed loop, as confirmed by the computed eigenvalues:
\begin{lstlisting}[style=matlabstyle]
eig_closedloop=HmqNEig(A-B*K_final,20,T,"fundamental")
\end{lstlisting}

\begin{lstlisting}[style=interpretercode]
>> eig_closedloop =
  -2.3234 + 0.0000i
  -3.4466 - 0.0000i
\end{lstlisting}

\subsection{LMI-Based Synthesis}
More importantly, Toeplitz-block LMIs for $H_2$ or $H_\infty$ control design \cite{vernerey_tblmi_2025} can also be constructed directly using the toolbox. In contrast, their time-domain counterparts take the form of differential LMIs, for which no systematic solution methods currently exist. For illustration purpose, consider the equivalent LMI formulation of the LQR problem~\cite{willemsLeastSquaresStationary1971,vernerey_tblmi_2025}:

\begin{equation}
\begin{aligned}
\max_{P} \quad & \text{trace}(P_0) \\
\text{s.t.} \quad & \hm P \succ 0 \\
& \begin{bmatrix}
(\hm A-\hm N)^* \hm P + \hm P (\hm A-\hm N) + \hm Q & \hm P \hm B \\
\hm B^* \hm P & \hm R
\end{bmatrix} \succeq 0.
\end{aligned}
\end{equation}
This LMI is implemented as follows, with full compatibility with YALMIP and the Mosek solver:


\begin{lstlisting}[style=matlabstyle]
% TBLMI LQR Optimization Problem
hlmi = 20; %truncation order for LMI
hP = 10; %truncation order for P in LMI
ht = 10; %truncation order for A in LMI
% Define decision variable (periodic P via harmonics)
P = PhasorArray.ndsdpvar(2, 2, hP);
% Formulate LMI components
LMIvar11 = trunc(A,ht).'*P + P*trunc(A,ht)+ P.d(T) + Q;
LMIvar12 = P*B;
LMIvar22 = R;
% Convert to Toeplitz-Block form
P_tb = P.T_tb(hlmi); 
LMI_tb = [T_tb(LMIvar11, hlmi) , T_tb(LMIvar12, hlmi);
 T_tb(LMIvar12.', hlmi), T_tb(LMIvar22, hlmi)];
% Solve: maximize trace(P_0) subject to constraints
Constraints = [P_tb >= 0, LMI_tb >= 0];
Objective = -trace(P{:,:,0});%use P{i,j,k} to access kth phasor of P_{i,j}
sol = optimize(Constraints, Objective);
\end{lstlisting}
\begin{remark}
Note that to avoid introducing inconsistencies with the required Toeplitz structure, any product, say AB, appearing in an LMI must first be computed using their PhasorArray representation and then converted into Toeplitz form using \texttt{T\_tb(A*B,h)}.
 We recall that \texttt{T\_tb(A,h)*T\_tb(B,h)} is not Toeplitz and not equal to \texttt{T\_tb(A*B,h)}. Note also that the term $- \hm N^* \hm P - \hm P \hm N = \hm{T}(\dot P)$ is obtained by \texttt{T\_tb(P.d(T),hlmi)}. Finally, a detailed discussion on the selection of the truncation orders \texttt{hlmi}, \texttt{hP}, and \texttt{ht} can be found in \cite{vernerey_tblmi_2025}.
\end{remark}

\begin{figure}[ht!]
\centering
\begin{subfigure}{0.49\columnwidth}
\centering
\includegraphics[width=\linewidth]{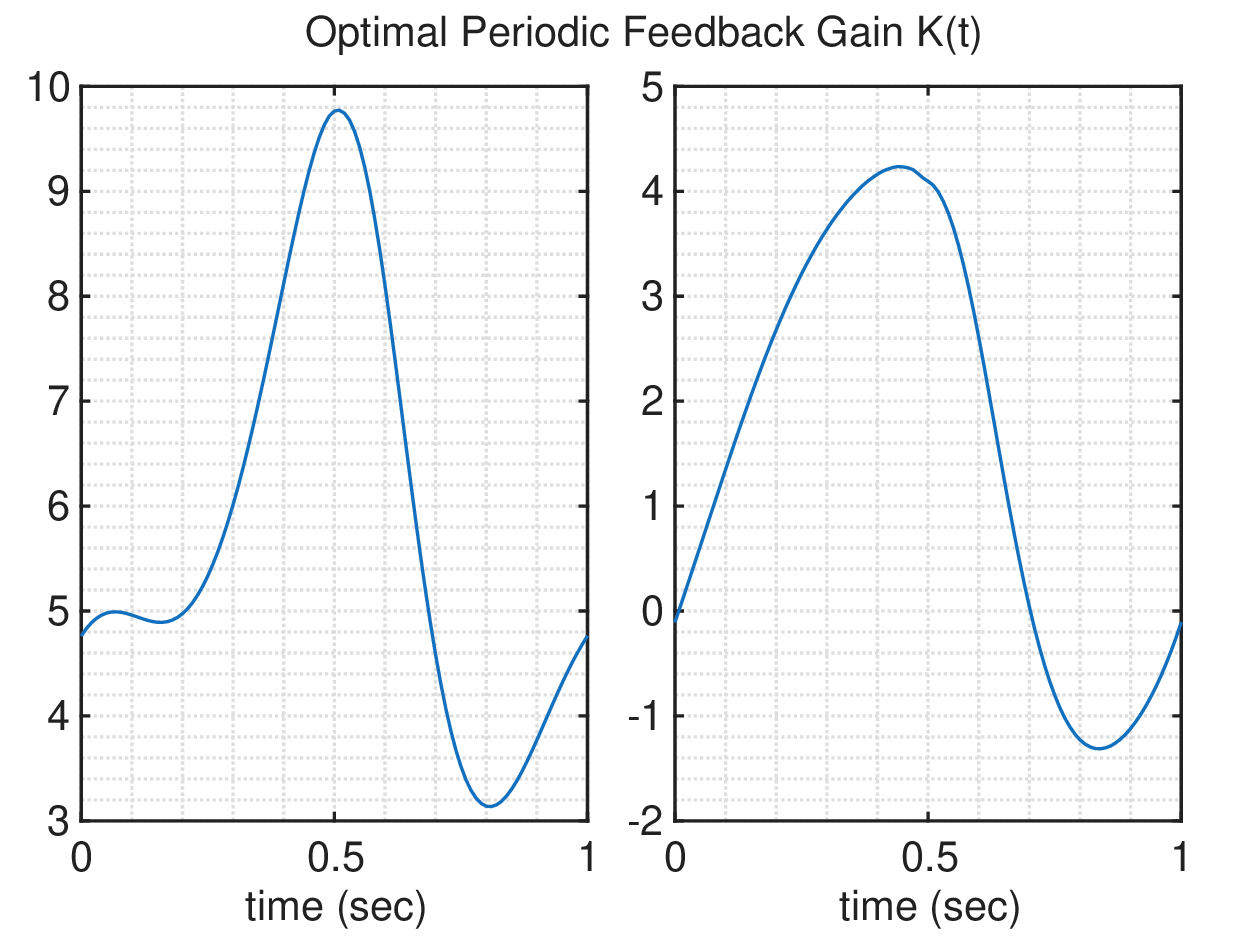}
\caption{Feedback gain $K(t)$}
\label{fig:RicK}
\end{subfigure}
\begin{subfigure}{0.49\columnwidth}
\centering
\includegraphics[width=\linewidth]{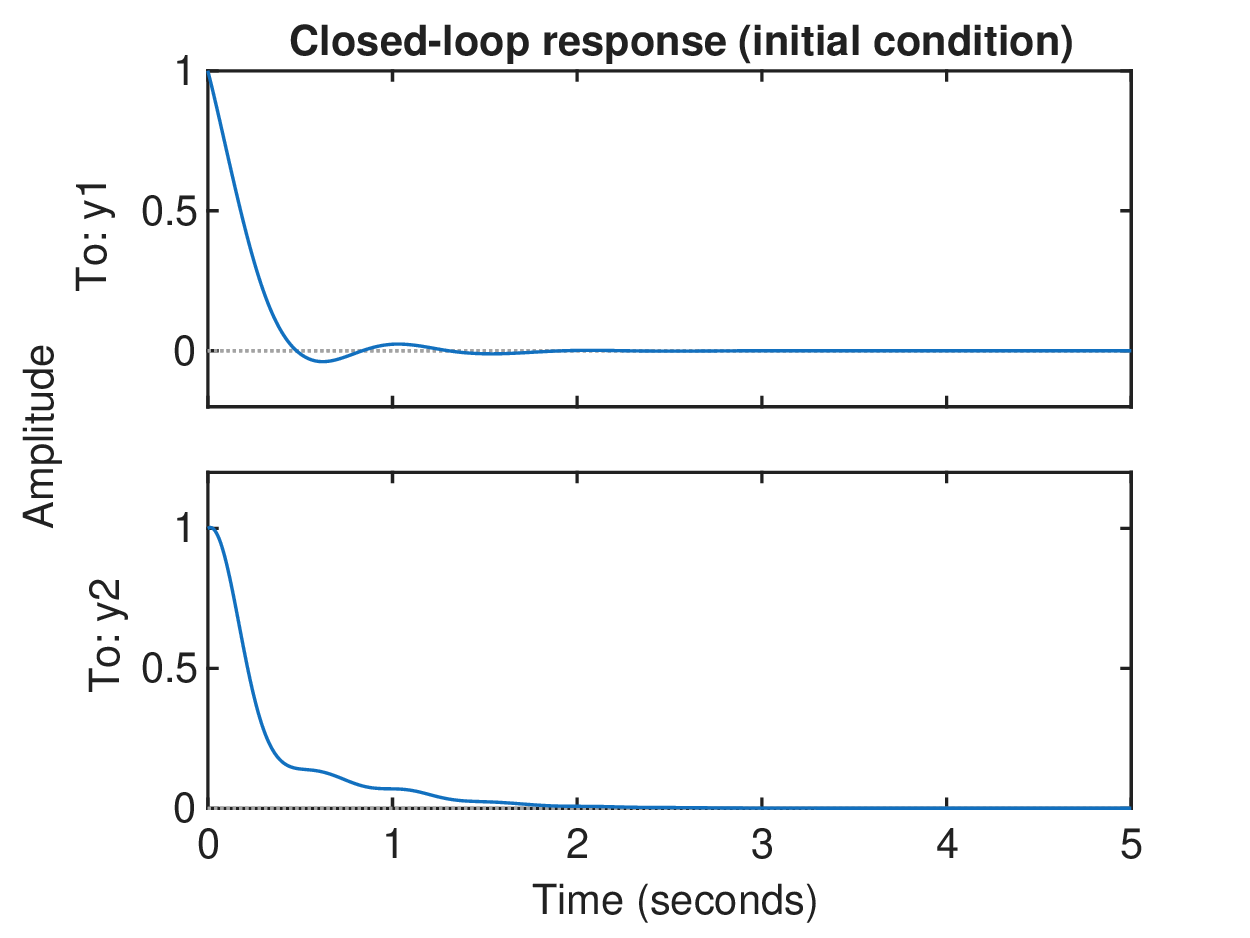}
\caption{Stabilized dynamics }
\label{fig:Sdet}
\end{subfigure}
\caption{Closed-loop response using \texttt{phasorSS}}
\end{figure}

Kleinman’s Riccati method and the TBLMI optimization problem yield comparable results. To illustrate this, Figures~\ref{fig:RicK} and~\ref{fig:Sdet} show the resulting periodic feedback gain $K(t)$ and the closed-loop response from an initial condition.

\begin{lstlisting}[style=matlabstyle]
Psol = sdpval(P); % Extract solution as PhasorArray
K=inv(R)*B'*Psol % Optimal feedback
% Compare solutions
figure
plot(Psol, T, 0:T/100:T);
hold on;
plot(S_final, T, 0:T/100:T)
figure
plot(K, T, 0:T/100:T)
\end{lstlisting}

\begin{figure}[ht!]
\centering
\includegraphics[width=0.6\columnwidth]{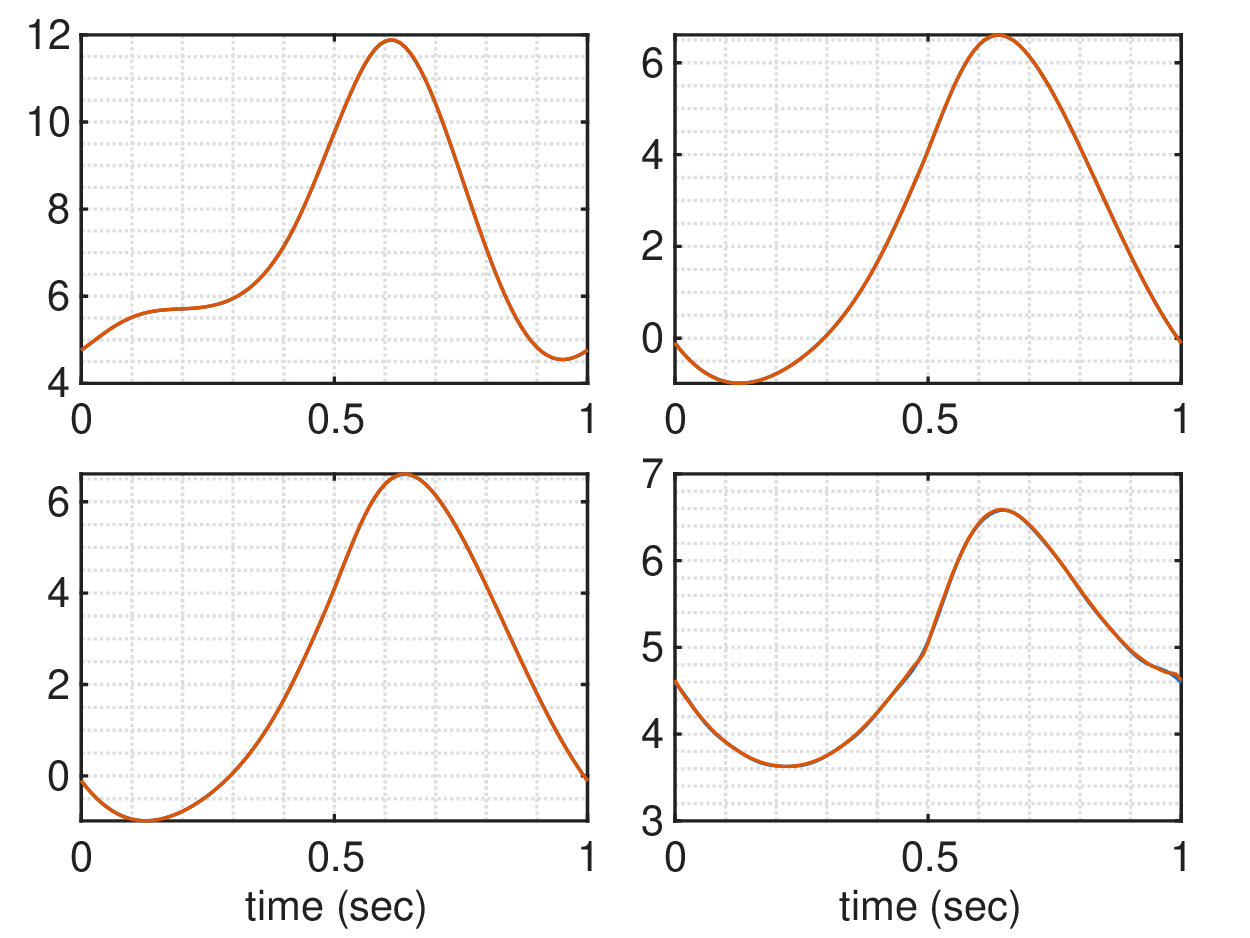}
\caption{Comparison of solution of Riccati $S(t)$ and LMI $P(t)$.}
\label{fig:riccati_vs_lmi}
\end{figure}

\subsection{System Simulation with PhasorSS}\label{sec:PhasorSS}

The \texttt{PhasorSS} class generalizes this framework to LTP systems and supports comprehensive system simulation, including step, initial-condition, and input–output responses.

\begin{lstlisting}[style=matlabstyle]
% Create LTP state-space system
C = PhasorArray.eye(2);
D = PhasorArray.zeros(2, 2);
sys = PhasorSS(A, B, C, D, T);
% Closed-loop system
sys_cl = feedback(sys, K_final);

figure
plot(sys_cl) %plot of matrix [A-BK,B;C,-DK]
\end{lstlisting}

\begin{lstlisting}[style=matlabstyle]
% Simulate initial condition response
x0 = [1; 1];
t_sim = 0:T/100:5*T;

figure
initial(sys_cl, x0, t_sim);
title('Closed-loop response (initial condition)')
figure
u= [1+PhasorArray.cos];
[y,t]=lsim(sys_cl,t_sim,u,x0);
plot(t,y);
title('Closed-loop response')
\end{lstlisting}

This demonstrates the complete workflow: from periodic matrices through stability analysis and controller synthesis, all the way to simulation and validation.

\section{Conclusion}






The PhasorArray Toolbox implements harmonic control methods by encapsulating periodic matrix operations within an object-oriented framework. Its 3D array structure, combined with overloaded operators, enables direct implementation of the theoretical results from~\cite{Blin2022NecessaryTime,riedinger_solving_2022,vernerey_tblmi_2025} using standard MATLAB syntax.

Key benefits include systematic handling of indexing and Toeplitz constructions behind the scenes, resulting in shorter and more readable code; a natural workflow from modeling to control synthesis without obscuring implementation details; and lowered entry barriers to harmonic-domain methods, facilitating broader adoption of these advanced techniques in the engineering community.


\bibliographystyle{IEEEtran}
\bibliography{references}

\end{document}